\documentclass[fleqn,twoside]{article}
\usepackage{espcrc2}

\newcommand{\be}{\begin{equation}}
\newcommand{\ee}{\end{equation}}
\newcommand{\bem}{\begin{displaymath}}
\newcommand{\eem}{\end{displaymath}}
\newcommand{\ba}{\begin{eqnarray}}
\newcommand{\ea}{\end{eqnarray}}
\newcommand{\re}[1]{(\ref{#1})}

\newcommand{\bM}{\breve{M}}

\newcommand{\bS}{\bar{S}} 
\newcommand{\bs}{\bar{{\cal S}}}

\newcommand{\1}{^{-1}}

\newcommand{\dg}{^{\dagger}}

\newcommand{\e}{\mbox{e}}

\newcommand{\f}{{\mbox{\scriptsize f}}}

\newcommand{\bG}{\bar{G}} 
\newcommand{\G}{{\cal B}} 
\newcommand{\ga}{\gamma_5}
\newcommand{\h}{\frac{1}{2}}
\newcommand{\Id}{\mbox{1\hspace{-.95mm}l}}   

\newcommand{\la}{\lambda}

\newcommand{\mo}[1]{^{(\mbox{\scriptsize #1})}}

\newcommand{\Pa}{{\cal P}} 
\newcommand{\bP}{\bar{P}} 

\newcommand{\s}{{\cal S}} 
\newcommand{\T}{{\cal T}} 
\newcommand{\Tr}{\mbox{Tr}} 
\newcommand{\bu}{\bar{u}} 
\newcommand{\U}{{\cal U}} 
\newcommand{\vp}{\varphi} 
\newcommand{\bw}{\bar{w}}
\newcommand{\W}{{\cal W}}

\newcommand{\AmS}{{\protect\the\textfont2
  A\kern-.1667em\lower.5ex\hbox{M}\kern-.125emS}}

\title{\vspace*{-10mm}
\raisebox{0.8cm}[0pt][0pt]{\makebox[0pt][l]{\parbox{16cm}{\normalsize%
\mbox{}\hfill HUB-EP-04/60}}}\\
Formulation of chiral gauge theories\thanks{
Talk presented at Lattice 2004, Fermilab, USA.}}

\author{Werner Kerler \address{Institut f\"ur Physik, 
        Humboldt-Universit\"at, D-12489 Berlin, Germany}%
}
       
\begin{document}

\begin{abstract}
We present a formulation of chiral gauge theories, which admits more general 
spectra of Dirac operators and reveals considerably more possibilities for 
the structure of the chiral projections. Our two forms of correlation 
functions both also apply in the presence of zero modes and for any value 
of the index. The decomposition of the total set of pairs of bases into 
equivalence classes is carefully analyzed. Transformation properties are 
derived.
\vspace{1pc}
\end{abstract}

\maketitle
\thispagestyle{empty}

\section{CHIRAL PROJECTIONS}

Starting from the basic structure of previous approaches to chiral gauge theories \cite{na93,lu98}
we have recently presented a generalization \cite{ke03} in which the
Dirac operator and the chiral projections have been considered as functions 
of a certain unitary and $\ga$-Hermitian operator. Here we avoid the 
restrictions introduced by referring to such an operator by determining the 
possible structures of the chiral projections for given Dirac operator $D$.

For operators satisfying $[D\dg,D]=0$ and $D\dg=\ga D\ga$ we have
the spectral representation 
\be
D=\sum_j\hat{\la}_j(P_j^++P_j^-) 
+\sum_{k}(\la_kP_k\mo{I}+\la_k^*P_k\mo{II})
\label{specd}
\ee
with $\mbox{Im }\hat{\la}_j=0$ and $\mbox{Im }\la_k>0$ and 
where 
$\ga P_j^{\pm}=P_j^{\pm}\ga=\pm P_j^{\pm}$ and $\ga P_k\mo{I}=P_k\mo{II}\ga$.
Since $\Tr(\ga\Id)=\Tr(\ga P_k\mo{I})=\Tr(\ga P_k\mo{II})=0$
we get for $N_j^{\pm}=\Tr\,P_j^{\pm}$
\be
\sum_j(N_j^+-N_j^-)=0.
\ee
Associating $j=0$ to zero modes the index of $D$ is 
given by $I=N_0^{+}-N_0^{-}$. 

In contrast to the Dirac operators considered previously those in \re{specd}
are no longer restricted to one real eigenvalue in addition to zero and also
admit more general complex ones. They have nevertheless appropriate 
realizations which also allow numerical evaluation \cite{ke04}. 

For the chiral projections $P_-$ and $\bP_+$ the fundamental relation
\be
\bP_+D=DP_-
\label{BAS}
\ee 
is required. Then because of $[P_-,DD\dg]=[\bP_+,DD\dg]=0$ we obtain 
the decomposition
\be
P_-=\sum_jP_j^{\rm X}+\sum_kP_k^{\rm R},\;\;
\bP_+=\sum_j\bP_j^{\rm X}+\sum_k\bP_k^{\rm R}
\label{DEC}
\ee
in which the projections $P_k^{\rm R}$ and $\bP_k^{\rm R}$ are given by
\ba
P_k^{\rm R}=c_kP_k\mo{I}+(1-c_k)P_k\mo{II}\hspace{25mm}\nonumber\\
-\sqrt{c_k(1-c_k)}\ga(\e^{i\vp_k}P_k\mo{I}+\e^{-i\vp_k}P_k\mo{II}),\\
\bP_k^{\rm R}=c_kP_k\mo{I}+(1-c_k)P_k\mo{II}\hspace{25mm}\nonumber\\
+\sqrt{c_k(1-c_k)}\ga\big(\e^{-i\bar{\vp}_k}P_k\mo{I}+\e^{i\bar{\vp}_k}
P_k\mo{II}\big),
\ea
where $0\le c_k\le1$, $\;\;\e^{i(\vp_k+\bar{\vp}_k-2\alpha_k)}=-1$,
$\;\;\e^{i\alpha_k}=\la_k/|\la_k|$ and 
\be
\Tr\,P_k^{\rm R}=\Tr\,\bP_k^{\rm R}=\Tr\,P_k\mo{I}=\Tr\,P_k\mo{II}
=\,:\tilde{N}_k.
\ee 
For the other projections, with $\bar{N}-N=I$ for $\bar{N}=\Tr\,\bP_+$ 
and $N=\Tr\,P_-$, we get
\be
\bP_0^{\rm X}=P_0^+,\qquad P_0^{\rm X}=P_0^-,
\ee
and have for $j\ne0$ the two possibilities  
\be
\bP_j^{\rm X}=P_j^{\rm X}=P_j^+\quad\textbf{ or }\quad \bP_j^{\rm X}=
P_j^{\rm X}=P_j^-.
\label{PoP}
\ee

With these relations for the chiral projections we see that, introducing 
$\Tr\,\Id=2d$, we have 
\be
\bar{N}=d,\;N=d-I\;\;\textbf{ or }\;\;
\bar{N}=d+I,\;N=d
\ee
for the two choices in \re{PoP}, respectively, and that
\ba
N=N_0^-+L,\qquad\bar{N}=N_0^++L,\nonumber\\
L=\sum_{j\ne0}N_j^{\pm}+\sum_k\tilde{N}_k
\label{LL}
\ea
holds, where $\pm$ refers to such two choices.

\section{CORRELATION FUNCTIONS}

Non-vanishing fermionic correlation functions are given by
\be
\langle\psi_{\sigma_{r+1}}\ldots\psi_{\sigma_N}\bar{\psi}_{\bar{\sigma}_{r+1}}
\ldots\bar{\psi}_{\bar{\sigma}_{\bar{N}}}\rangle_{\f}=
\label{COR}
\ee
\bem
\frac{1}{r!}\sum_{\bar{\sigma}_1\ldots\bar{\sigma}_r}\sum_{\sigma_1,\ldots,
\sigma_r}\bar{\Upsilon}_{\bar{\sigma}_1\ldots\bar{\sigma}_{\bar{N}}}^*
\Upsilon_{\sigma_1\ldots\sigma_N}D_{\bar{\sigma}_1\sigma_1}\ldots
D_{\bar{\sigma}_r\sigma_r}\nonumber
\eem
with the alternating multilinear forms
\ba
\Upsilon_{\sigma_1\ldots\sigma_N}=\sum_{i_1,\ldots,i_N=1}^N\epsilon_{i_1, 
\ldots,i_N}u_{\sigma_{1}i_{1}}\ldots u_{\sigma_Ni_N},\\\bar{\Upsilon}_{
\bar{\sigma}_1\ldots{\bar{\sigma}_{\bar{N}}}}=\sum_{j_1,\ldots
j_{\bar{N}}=1}^{\bar{N}}\epsilon_{j_1,\ldots,j_{\bar{N}}}\bar{u}_{\bar{
\sigma}_{1}j_{1}}\ldots\bar{u}_{\bar{\sigma}_{\bar{N}}j_{\bar{N}}},
\ea
in which the bases $\bu_{\bar{\sigma}j}$ and $u_{\sigma i}$ satisfy
\be
P_-=uu\dg,\; u\dg u=\Id_{\rm w},\;\;\bP_+=\bu\bu\dg,\;\bu\dg\bu=
\Id_{\rm\bw}.
\label{uu}
\ee 

While $P_-$ and $\bP_+$ are invariant under unitary basis transformations 
$u^{(S)}=uS$, $\bu^{(\bar{S})}=\bu\bar{S}$, the forms
$\Upsilon_{\sigma_1\ldots\sigma_N}$ and $\bar{\Upsilon}_{\bar{\sigma}_1
\ldots{\bar{\sigma}_{\bar{N}}}}$ get multiplied by $\det_{\rm w}S$ and 
$\det_{\rm \bw}\bar{S}$, respectively. Therefore, in order that all general 
correlation functions remain invariant we have to impose the condition
\be
{\det}_{\rm w}S\cdot{\det}_{\rm\bw}\bar{S}\dg=1.
\label{dd}
\ee
Without this condition all bases related to a chiral projection are
connected by unitary transformations. With it the total set of pairs of 
bases $u$ and $\bu$ decomposes into equivalence classes of which one is 
to be chosen to describe physics. Different equivalence classes are 
related by pairs of basis transformations with
\be
{\det}_{\rm w}S\cdot{\det}_{\rm\bw}\bar{S}\dg=\e^{i\Theta},\quad\Theta\ne0.
\label{INE}
\ee
The phase factor $\e^{i\Theta}$ determines how the results
of the respective formulations of the theory differ. 

The relations obtained for the chiral projections imply ones for the bases,
too. Thus from 
\be
\bP_k^{\rm R}=|\la_k|^{-2}DP_k^{\rm R}D\dg
\ee
putting $P_k^{\rm R}=\sum_{l=1}^{\tilde{N}_k}u_l^{[k]}u_l^{[k]\dag}$ we get 
\be
\bu_l^{[k]}=\e^{-i\Theta_k}|\la_k|\1Du_l^{[k]}
\label{M1}
\ee
with phases $\Theta_k$ so that $\bP_k^{\rm R}=\sum_{l=1}
^{\tilde{N}_k}\bu_l^{[k]}\bu_l^{[k]\dag}$.
For $P_j^{\pm}=\sum_{l=1}^{N_j^{\pm}}u_l^{\pm[j]}u_l^{\pm[j]\dag}
=\sum_{l=1}^{N_j^{\pm}}\bar{u}_l^{\pm[j]}\bar{u}_l^{\pm[j]\dag}$
where $j\ne0$ we have with phases $\Theta_j^{\pm}$ 
\be
\bu_l^{\pm[j]}=\e^{-i\Theta_j^{\pm}}|\hat{\la}_j|\1Du_l^{\pm[j]}.
\label{M2}
\ee

\hspace{0mm}From \re{M1} and \re{M2} it becomes obvious that the 
$L\times L$ submatrix $\bM$ of $\bu\dg Du$, which occurs according 
to \re{LL}, has the eigenvalues 
\be
\e^{i\Theta_k}|\la_k|,\qquad\quad\e^{i\Theta_j^{\pm}}|\hat{\la}_j|,  
\ee
with multiplicities $\tilde{N}_k$ and $N_j^{\pm}$, respectively. 
Using $\bM$ and introducing $P_0^-=\sum_{l=L+1}^Nu_lu_l\dg$ and 
$P_0^+=\sum_{l=L+1}^{\bar{N}}\bu_l\bu_l\dg$ for the zero mode part we 
find for the correlation functions the form
\be
\langle\psi_{\sigma_{r+1}}\ldots\psi_{\sigma_N}\bar{\psi}_{\bar{\sigma}_{r+1}}
\ldots\bar{\psi}_{\bar{\sigma}_{\bar{N}}}\rangle_{\f}=
\ee
\bem
\sum_{\sigma_{r+1}',\ldots,\sigma_N'}\epsilon\,_{\sigma_{r+1}
\ldots\sigma_N}^{\sigma_{r+1}'\ldots\sigma_N'}\;\sum_{\bar{\sigma}_{r+1}',
\ldots,\bar{\sigma}_{\bar{N}}'}\epsilon\,_{\bar{\sigma}_{r+1}\ldots
\bar{\sigma}_{\bar{N}}}^{\bar{\sigma}_{r+1}'\ldots\bar{\sigma}_{\bar{N}}'}\;
\;{\textstyle\frac{1}{(L-r)!}}
\eem
\bem
{\cal G}_{\sigma_{r+1}'\bar{\sigma}_{r+1}'}\ldots{\cal G}_{\sigma_L'\bar{
\sigma}_L'}\quad
\e^{-i\theta_{\rm z}^-}\;u_{\sigma_{L+1},L+1}\ldots u_{\sigma_NN}
\eem
\bem
\e^{i\theta_{\rm z}^+}\;\bu_{L+1,\bar{\sigma}_{L+1}}\dg\ldots
\bu_{\bar{N}\bar{\sigma}_{\bar{N}}}\dg\;\;{\det}_L\bM,
\eem
with ${\cal G}=\breve{P}_-\breve{D}\1\breve{\bP}_+$, where $\breve{D}$, 
$\breve{P}_-$, $\breve{\bP}_+$ are the restrictions of the operators $D$, 
$P_-$, $\bP_+$ to the subspace on which $\Id-P_0^+-P_0^-$ projects. 
With $\theta_{\rm z}^+$ and $\theta_{\rm z}^-$ being related to the 
zero-mode part, the
equivalence class of pairs of bases is characterized by the value of
\be
\sum_kN_k\Theta_k+\sum_{j\ne0}N_j^{\pm}\Theta_j^{\pm}+\theta_{\rm z}^+-
\theta_{\rm z}^-.
\ee

\section{GAUGE TRANSFORMATIONS}

Conditions \re{uu} and \re{dd} determine the equivalence class of pairs of 
bases $uS$, $\bu\bS$. Gauge transformations $P_-'=\T P_-\T\dg$, $\bP_+'=
\T\bP_+\T\dg$ for $[\T,P_-]\ne0$, $[\T,\bP_+]\ne0$ imply that 
the transformed equivalence class is given by
\be
u'S'=\T uS\s,\qquad\bu'\bS'=\T\bu\bS\bs,
\label{TRA}
\ee
where $u'$, $\bu'$, $S'$, $\bS'$ satisfy the transformed conditions \re{uu} 
and \re{dd}, and where the unitary transformations $\s(\T,\U)$ and 
$\bs(\T,\U)$ with ${\det}_{\rm w}\s(\Id,\U)\big({\det}_{\rm\bw}\bs(\Id,\U)
\big)^*=1$ are introduced for full generality.  Insertion of \re{TRA} 
into \re{COR} gives for the correlation functions 
\be
\langle\psi_{\sigma_1'}'\ldots\psi_{\sigma_R'}'\bar{\psi}_{\bar{\sigma}_1'}'
\ldots\bar{\psi}_{\bar{\sigma}_{\bar{R}}'}'\rangle_{\f}'=
\label{CORG}
\ee
\bem\e^{i\vartheta_{\T}}
\sum_{\sigma_1,\ldots,\sigma_R}\sum_{\bar{\sigma}_1,\ldots,\bar{\sigma}_{
\bar{R}}}\T_{\sigma_1'\sigma_1}\ldots\T_{\sigma_R'\sigma_R}
\eem
\bem
\hspace*{8mm}\langle\psi_{\sigma_1}\ldots\psi_{\sigma_R}\bar{\psi}_{\bar{
\sigma}_1}\ldots\bar{\psi}_{\bar{\sigma}_{\bar{R}}}\rangle_{\f}\quad
\T_{\bar{\sigma}_1\bar{\sigma}_1'}\dg\ldots
\T_{\bar{\sigma}_{\bar{R}}\bar{\sigma}_{\bar{R}}'}\dg.
\eem
In this relation the factor 
\be
\e^{i\vartheta_{\T}}={\det}_{\rm w}\s\cdot{\det}_{\rm\bw}\bs\dg
\ee 
for $\vartheta_{\T}\ne0\,$ has just the form met in \re{INE} for the 
transformations to inequivalent subsets of pairs of bases. Thus to prevent
arbitrary switching to different equivalence classes the condition
$\vartheta_{\T}=0\,$ is to be imposed. That this is to be done follows
\cite{ke04}, on the other hand, also from the covariance requirement for 
the current in Ref.~\cite{lu98}.

In the special case $[\T,P_-]\ne0$, $[\T,\bP_+]=0$ the equivalence class
can be represented \cite{ke04} by pairs $uS$, $\bu_{\rm c}\bS_{\rm c}$ 
where $\bu_{\rm c}$ and $\bS_{\rm c}$ are independent of the gauge field 
so that instead of \re{TRA} we have
\be
u'S'=\T uS\s,\qquad\bu_{\rm c}\bS_{\rm c}=\mbox{const}.
\label{TRA1}
\ee
Because of $[\T,\bP_+]=0$ it is now possible to rewrite 
$\bu_{\rm c}=\T\bu_{\rm c}\hat{S}_{\T}$. With this and \re{TRA1} we get
again the form \re{CORG}, however, with
\be
\e^{i\vartheta_{\T}}={\det}_{\rm w}\s\cdot{\det}_{\rm\bw}\hat{S}_{\T}\dg.
\ee
Here ${\det}_{\rm w}\s=1$ remains to be required to prevent arbitrary 
switching to different equivalence classes. For the factor 
${\det}_{\rm\bw}\hat{S}_{\T}\dg$ with $\T=\exp(\G)$ we obtain the
constant result
\be
{\det}_{\rm\bw}\hat{S}_{\T}\dg =\exp({\textstyle\h}\Tr\,\G).
\ee 

It should be noted that in the continuum limit certain compensations of
terms present on the lattice disappear so that one arrives just at the 
usual features of continuum perturbation theory \cite{ke03,ke04}.

\section{CP TRANSFORMATIONS}

For CP transformations of the chiral projections we have 
\ba
P_-^{\rm CP}(\U^{\rm CP})=\W\bP_+^{\rm T}(\U)\W\dg,\\ 
\bP_+^{\rm CP}(\U^{\rm CP})=\W P_-^{\rm T}(\U)\W\dg,\,
\ea
with $\W=\Pa\gamma_4C\dg$,
$\Pa_{n'n}=\delta^4_{n'\tilde{n}}$, $U_{4n}^{\rm CP}=U_{4\tilde{n}}^*$ 
and $U_{kn}^{\rm CP}=U_{k,\tilde{n}-\hat{k}}^*$ for $k=1,2,3$, where 
$\tilde{n}=(-\vec{n},n_4)$.
Writing $P_-(\U)$ and $\bP_+(\U)$ in the form 
\be
P_-=\h(\Id-\ga G),\quad\bP_+=\h(\Id+\bG\ga)
\ee
we get for $P_-^{\rm CP}(\U^{\rm CP})$ and $\bP_+^{\rm CP}(\U^{\rm CP})$ 
\be
P_-^{\rm CP}=\h\big(\Id-\ga\bG\big),
\quad\bP_+^{\rm CP}=\h\big(\Id+G\ga\big).
\ee
Obviously the transformed projections differ by an interchange of $G$ and 
$\bG$, in which context it is to be noted that generally $\bG\ne G$ 
holds \cite{ke04}.

With the conditions \re{uu} and \re{dd} satisfied by $u$, $\bu$, $S$, 
$\bS$ as well as by $u^{\rm CP}$, $\bu^{\rm CP}$, $S^{\rm CP}$, 
$\bS^{\rm CP}$, the equivalence class of pairs of bases transforms as
\be
u^{\rm CP}S^{\rm CP}=\W\bu^*\bS^*S_{\zeta},\;\;\bu^{\rm CP}
\bS^{\rm CP}=\W u^*S^*\bS_{\zeta}
\ee
where the unitary transformations $S_{\zeta}$ and $\bS_{\zeta}$ are 
introduced for full generality. Inserting this into \re{COR} we get for 
the correlation functions
\be
\langle\psi_{\sigma_1'}^{\rm CP}\ldots\psi_{\sigma_R'}^{\rm CP}\bar{\psi}_{
\bar{\sigma}_1'}^{\rm CP}\ldots\bar{\psi}_{\bar{\sigma}_{\bar{R}}'}^{\rm CP}
\rangle_{\f}^{\rm CP}=
\ee
\bem
\e^{i\vartheta_{\rm CP}}
\sum_{\sigma_1,\ldots,\sigma_R}\sum_{\bar{\sigma}_1,\ldots,\bar{\sigma}_{
\bar{R}}}\W_{\bar{\sigma}_1\bar{\sigma}_1'}\dg\ldots\\\ldots
\W_{\bar{\sigma}_{\bar{R}}\bar{\sigma}_{\bar{R}}'}\dg
\eem
\bem
\hspace*{5mm}\langle\psi_{\bar{\sigma}_1}\ldots\psi_{\bar{\sigma}_{\bar{R}}}
\bar{\psi}_{\sigma_1}\ldots\bar{\psi}_{\sigma_R}
\rangle_{\f}\quad\W_{\sigma_1'\sigma_1}\ldots\W_{\sigma_R'\sigma_R}.
\eem
Here the factor
\be
\e^{i\vartheta_{\rm CP}}={\det}_{\rm\bw}S_{\zeta}\cdot
{\det}_{\rm w}\bar{S}_{\zeta}\dg
\label{INC}
\ee
is subject to the condition that repetition of the transformation must 
always lead back, which is satisfied by restricting $S_{\zeta}$ and 
$\bar{S}_{\zeta}$ to choices for which $\vartheta_{\rm CP}$ is a universal 
constant. Then the resulting factor gets irrelevant in full correlation 
functions so that one may put $\vartheta_{\rm CP}=0$.

\vspace*{4mm}
I wish to thank Michael M\"uller-Preussker and his group for their kind
hospitality.


\begin{thebibliography}{9}
\bibitem{na93}  R. Narayanan, H. Neuberger, 
              Phys. Rev. Lett. 71 (1993) 3251;
              Nucl. Phys. B412 (1994) 574; 
              Nucl. Phys. B443 (1995) 305. 
\bibitem{lu98}  M. L\"uscher,
              Nucl. Phys. B549 (1999) 295; 
              Nucl. Phys. B568 (2000) 162. 
\bibitem{ke03}  W. Kerler, Nucl. Phys. B680 (2004) 51. 
\bibitem{ke04}  W. Kerler, hep-lat/0402011. 
\end{thebibliography}
\end{document}